\documentclass[12pt]{article}
\usepackage{amssymb}

\def\e{\begin{equation}} 
\def\f{\end{equation}} 
\def\ea{\begin{eqnarray}} 
\def\fa{\end{eqnarray}} 

\def\##1{{\bf #1\mit}}
\def\%#1{{\mbox{\boldmath $#1$}}}
\def\=#1{{\overline{\overline{\mathsf #1}}}}

\def\/{\over}
\def\*{^{\displaystyle*}}

\def\.{\cdot}
\def\x{\times}
\def\:{\over}
\def\oo{\infty}

\def\D{\nabla}
\def\d{\partial}

\def\ra{\rightarrow}

\def\l#1{\label{eq:#1}}
\def\r#1{(\ref{eq:#1})}
\def\am{\left(\begin{array}{c}}
\def\amm{\left(\begin{array}{cc}}
\def\a{\end{array}\right)}

\def\B{\beta}

\def\E{\epsilon}

\def\h{\eta}

\def\la{\lambda}

\def\M{\mu}
\def\o{\omega}

\begin{document}

\title{Realization of the D'B' Boundary Condition}
\author{I.V. Lindell$^1$, A. Sihvola$^1$, L. Bergamin$^2$, A. Favaro$^3$\\
${}^1$ Aalto University, Espoo, Finland\\ ${}^2$ KB\&P GmbH, Bern, Switzerland,\\ ${}^3$ Imperial College London, United Kingdom}%

\maketitle

\begin{abstract}
In this paper we find a realization for the D'B'-boundary conditions, which imposes vanishing normal derivatives of the normal components of the D and B fields. The implementation of the DB boundary, requiring vanishing normal components of D and B, is known. It is shown that the realization of the D'B' boundary can be based on a layer of suitable metamaterial, called the wave-guiding quarter-wave transformer, which transforms the DB boundary to the D'B' boundary. In an appendix, the mixed-impedance boundary, which is a generalization of both DB and D'B' boundaries, is shown to transform to another mixed-impedance boundary through the same transformer. 
\end{abstract}

\section{Introduction}

Electromagnetic boundary-value problems are normally defined in terms of impedance-boundary conditions. In turn, these involve linear relations between electric and magnetic field components tangential to the boundary surface. Denoting by $\#n$ the unit vector normal to the boundary, the general form for such conditions can be written as \cite{Methods}
\e \#n\x\#E = \=Z_s\.\#H, \ \ \ {\rm or}\ \ \ \#n\x\#H = \=Y_s\.\#E=0, \l{imp}\f
where $\=Z_s$ and $\=Y_s$ are the surface-impedance and surface-admittance dyadics, satisfying $\#n\.\=Z_s=\=Z_s\.\#n=0$ and $\#n\.\=Y_s=\=Y_s\.\#n=0$. Thus, the normal components of the $\#E$ and $\#H$ vectors do not play any role in \r{imp}. Typical examples of impedance-boundary conditions are the PEC and PMC conditions, $ \#n\x\#E=0$ and $ \#n\x\#H=0,$ corresponding to the respective cases $\=Z_s=0$ and $\=Y_s=0$. It is alternatively possible to define boundary conditions in terms of the normal components of the fields. In fact, conditions of the form
\e \#n\.\#D=0,\ \ \ {\rm and}\ \ \ \#n\.\#B=0, \l{DB}\f
originally introduced in \cite{Rumsey}, have been shown to yield unique solutions to boundary-value problems \cite{Yee,Kress}. 

The conditions \r{DB} have recently proved important in constructing electromagnetic cloaking structures \cite{Kong,Yaghjian,Weder}. The conditions \r{DB} have subsequently been dubbed  DB-boundary conditions \cite{Hongkong,DB}. Other boundary conditions involving the normal components and/or the normal derivatives of normal components of the fields were introduced in \cite{AP10}; in particular the requirements
\e \D\.(\#n\#n\.\#D)=0,\ \ \ \ {\rm and}\ \ \ \ \D\.(\#n\#n\.\#B)=0 \l{D'B'}\f
were named D'B'-boundary conditions. It has been shown that objects with certain symmetry properties and defined by either DB or D'B' boundary conditions have zero backscattering, i.e., they cannot be seen by radar \cite{AP09}. 

All boundary conditions defined above are mathematical concepts. From the practical point of view one faces the problem of realizing them in terms of physical structures, as precisely as possible. It is known that the PEC boundary corresponds to an interface of an ideally conducting material, which can be approximated by metals. In \cite{Rumsey} it was shown that the DB boundary can be implemented by an interface of an anisotropic medium, whose normal permittivity and permeability parameters become zero. Other possibilities have also been suggested. Up till now it has been a problem to find a realization for the D'B' boundary. The problem will be addressed in this paper by assuming an isotropic half space $z<0$ with planar boundary $z=0$, at which the D'B' conditions \r{D'B'} have the form
\e \d_zD_z=0,\ \ \ \ \ \d_zB_z=0. \l{D'B'1}\f

The problem is handled in terms of a plane wave of fixed frequency incident to the boundary. It is obvious that, if the D'B' conditions are satisfied by an arbitrary plane wave, being linear, the conditions are satisfied by a field consisting of a sum or an integral of plane waves, in fact, by any electromagnetic field outside its sources. 

As a guiding principle in the realization we use the knowledge that the eigenpolarizations for the plane wave reflecting from the D'B' plane, as well as from the DB plane, are the TE and TM polarizations with respect to the normal of the plane. In fact, one can show that an incident TE wave is reflected as a TE wave and an incident TM wave is reflected as a TM wave. It is also known that for these eigenpolarizations a DB and D'B' plane can be replaced by a corresponding PEC and PMC plane, according to the following table \cite{AP10}
\begin{table}[ht]
	\centering
		\begin{tabular}{|c|c|c|}
		\hline
			 & TE & TM\\
			 \hline
			 DB & PEC & PMC\\
			 D'B' & PMC & PEC\\ 
			  \hline			
		\end{tabular}
	\caption{Boundary conditions involving normal field components can be replaced by effective PEC and PMC conditions for fields with TE and TM polarizations.}
	\label{tab:BC}
\end{table}

Since it is known that the eigenwaves in a uniaxially anisotropic medium are also TE and TM polarized \cite{Chen}, it appears natural to study the possibility of realization in terms of such a medium.

\section{Plane-wave in uniaxial anisotropic medium}

Let us consider the uniaxial anisotropic medium defined by the medium equations
\ea \#D &=& \E_t\#E_t + \E_z\#u_zE_z,\\
 \#B &=& \M_t\#H_t + \M_z\#u_zH_z,\fa
where $\#E_t$ and $\#H_t$ are components transverse to the $z$ axis. Assuming a plane-wave in the medium with fields of the form
\ea \#E(\#r) &=& \#E e^{-jk_xx}e^{-j\B z},\\
 \#H(\#r) &=& \#H e^{-jk_xx}e^{-j\B z}, \fa
the Maxwell equations can be written as
\ea (k_x\#u_x + \B\#u_z)\x\#E &=& \o\M_t\#H_t + \o\M_z\#u_z H_z, \\  
 (k_x\#u_x + \B\#u_z)\x\#H &=& -\o\E_t\#E_t - \o\E_z\#u_z E_z. \fa
When expanding in component form, the equations split in two groups, one consisting of
\ea k_xE_y &=& \o\M_zH_z\\
    \B_{TE}H_x - k_xH_z &=& -\o\E_tE_y\\
    -\B_{TE}E_y &=& \o\M_tH_x, \fa
which corresponds to the TE component, and
\ea k_xH_y &=& -\o\E_zE_z\\
    \B_{TM}E_x - k_xE_z &=& \o\M_tH_y\\
    -\B_{TM}H_y &=& -\o\E_tE_x, \fa
which corresponds to the TM component.
     
Eliminating the transverse fields, two equations remain, one for $B_z=\M_zH_z$ and the other one for $D_z=\E_zE_z$,
\ea (\B_{TE}^2+ k_x^2(\M_t/\M_z) - k_t^2)B_z&=&0,\\
 (\B_{TM}^2+ k_x^2(\E_t/\E_z) - k_t^2)D_z&=&0, \fa
with 
\e k_t=\o \sqrt{\M_t\E_t}. \f
The propagation factors can be solved as
\e \B_{TE} = \sqrt{k_t^2 - k_x^2(\M_t/\M_z)}, \l{BTE}\f
\e \B_{TM} = \sqrt{k_t^2 - k_x^2(\E_t/\E_z)}. \l{BTM} \f

Since all fields have the same $x$ dependence, $\exp(-jk_xx)$, we may omit it, and consider the axial fields consisting of waves propagating in the two directions $\pm \#u_z$ as
\ea D_z(z) &=& D_+ e^{-j\B_{TM} z} + D_-e^{j\B_{TM} z}, \\ 
B_z(z) &=& B_+ e^{-j\B_{TE} z} + B_-e^{j\B_{TE} z}. \fa
The transverse fields can then be expressed as
\ea E_y(z)&=& \frac{\o}{k_x}(B_+ e^{-j\B_{TE} z} + B_-e^{j\B_{TE} z}), \\
H_x(z) &=& -\frac{\B_{TE}}{k_x\M_t}(B_+ e^{-j\B_{TE} z} - B_-e^{j\B_{TE} z}), \\
H_y(z) &=& -\frac{\o}{k_x}(D_+ e^{-j\B_{TM} z} + D_-e^{j\B_{TM} z}),\\
E_x(z) &=& -\frac{\B_{TM}}{k_x\E_t}(D_+ e^{-j\B_{TM} z} - D_-e^{j\B_{TM} z}). \fa

\section{Layer of wave-guiding medium}

After some analysis, one can show that, while the DB boundary can be realized by the interface of a uniaxial medium (one with zero axial parameters), there does not seem to exist a corresponding medium realizing the D'B' boundary. The next best choice would be to use a layer of uniaxial medium, extending from $z=0$ to $z=d$, as a transformer, for the purpose of transforming a DB boundary at $z=d$ to a D'B' boundary at $z=0$. Because the boundary conditions must be independent of the parameter $k_x$, i.e., valid for all plane waves, let us assume that the axial parameters of the layer medium grow large as
\e \E_z/\E_t\ra\oo,\ \ \ \ \M_z/\M_t\ra\oo. \l{EzEt}\f
In this case the two propagation factors \r{BTE} and \r{BTM} are reduced to
\e \B_{TE}\ra k_t,\ \ \ \ \B_{TM}\ra k_t , \f
independent of the wavenumber $k_x$. Such a medium is an example of the wave-guiding medium introduced in \cite{AP07}. In a wave-guiding medium, a field with any amplitude distribution (in the plane transverse to the $z$ axis) will retain its distribution when propagating along the $z$ axis. This resembles propagation in a bunch of waveguides parallel to the $z$ axis with no interaction between the fields in the waveguides.

It has been shown that a layer of suitable wave-guiding medium (with transverse anisotropy allowed), terminated by a PEC plane at $z=d$ is able to produce any given boundary impedance dyadic $\=Z_s$ at the plane $z=0$ \cite{AP07}. Let us now study the possibility that the present wave-guiding medium (with transverse isotropy) can be used for transforming DB conditions to D'B' conditions. Requiring the DB conditions at $z=d$
\ea D_z(d) &=& D_+ e^{-jk_t d} + D_-e^{jk_t d}=0, \\ 
B_z(d) &=& B_+ e^{-jk_t d} + B_-e^{jk_t d}=0, \fa
we can solve for the coefficients $D_-$ and $B_-$ whence the normal fields at $z=0$ can be expressed as
\ea D_z(0) &=& 2jD_+ e^{-jk_t d}\sin(k_td), \\ 
B_z(0) &=& 2jB_+ e^{-jk_t d}\sin(k_t d). \fa
The transverse fields at the interface $z=0$ now become
\ea E_y(0)&=& \frac{2j\o}{k_x}B_+e^{-jk_t d}\sin(k_t d), \\
H_x(0) &=& -\frac{2k_t}{k_x\M_t}B_+e^{-jk_td}\cos(k_td), \\
H_y(0) &=& -\frac{2j\o}{k_x}D_+ e^{-jk_t d}\sin(k_t d),\\
E_x(0) &=& -\frac{2k_t}{k_x\E_t}D_+e^{-jk_td}\cos(k_td), \fa
and they are continuous across the interface.

\section{Realization of D'B' conditions}

Considering fields in the isotropic half space $z\leq0$ with parameters $\E_o,\M_o$, consisting of an incident and a reflected plane wave,
\e \#E(x,z) = e^{-jk_xx}(\#E^i e^{-jk_zz}+\#E^re^{jk_zz}), \f
we can match the tangential fields at the interface $z=0$ to those of the wave-guiding medium. Because the plane-wave fields have no sources, they satisfy
\ea \D\.\#D &=& \d_zD_z + \E_o\d_xE_x =0 \\
\D\.\#B &=& \d_zB_z + \E_o\d_xH_x =0, \fa
whence at the interface we have
\ea \d_zD_z(0) &=& jk_x\E_oE_x(0) \\ &=& -\frac{2jk_t\E_o}{\E_t}D_+e^{-jk_td}\cos(k_td), \\
\d_zB_z(0) &=& jk_x H_x(0) \\
&=& -\frac{2jk_t\M_o}{\M_t}B_+e^{-jk_td}\cos(k_td). \fa
It is now obvious that the D'B' conditions \r{D'B'1} can be enforced by choosing the thickness of the layer as
\e  d= \pi/2k_t. \f
Thus, the boundary with DB conditions at $z=d$ is transformed to a boundary with D'B' conditions at $z=0$. Defining the wavelength $\la$ in the wave-guiding medium by $k_t\la=2\pi$, whence $d=\la/4$, the structure can be called wave-guiding quarter-wave transformer. The transformation can be easily generalized to mixed boundary impedance conditions \cite{Mixed}. The equations are given as an Appendix.

Since the transverse parameters of the wave-guiding medium can be freely chosen, the transformer length $d$ can be made as small as one wishes by letting $\M_t\E_t$ grow large enough. Here we must, however, distinguish between the two different orders of magnitude involved in the axial and transverse medium quantities as is seen from \r{EzEt}. Recalling that the realization of the DB boundary (by vanishing axial parameters) may also lead to a thin sheet of material \cite{Kong,DB,ZAP}, the final realization of the D'B' boundary may theoretically be achieved by a thin double sheet. Because the boundary conditions are local, the same thin-sheet implementation apparently remains valid for smoothly curved boundary surfaces as well.

\section{Conclusion}
So far, the D'B' boundary conditions have not found a physical realization. The paper presents one solution for this problem. The implementation of the DB boundary is known in terms of the interface of a uniaxially anisotropic medium with zero axial parameters (ZAP medium). Here, it is suggested that the planar D'B' boundary is realized by transforming such a DB boundary by means of a wave-guiding quarter-wave transformer. Such a device is a quarter-wave slab of uniaxial medium with infinitely large axial parameters. Thus, the open question whether the D'B' boundary is just a mathematical artefact, without a physical counterpart, has been answered. In an appendix the quarter-wave transformer is shown to transform a set of more general mixed-impedance boundary conditions to a similar form with transformed parameters.

\section{Appendix: Mixed boundary conditions}

The conditions for the DB and D'B' boundaries have been extended to more general forms in \cite{AWPL09,Mixed}. In the form 
\e jk\h B_z -Z_{TE}\d_z B_z=0, \l{mixTE}\f
\e jk Z_{TM} D_z - \h\d_zD_z=0, \l{mixTM}\f
they include the DB conditions ($Z_{TE}=0$, $Z_{TM}=\oo$) and the D'B' conditions ($Z_{TM}=0$, $Z_{TE}=\oo$) as special cases. \r{mixTE} and \r{mixTM} were called mixed-impedance boundary conditions in \cite{Mixed} because the boundary is seen as the surface impedance $Z_{TE}$ by a TE field and as the surface impedance $Z_{TM}$ by a TM field. 

Assuming that \r{mixTE} and \r{mixTM} are valid at $z=d=\pi/2k_t$, the conditions for the field are
$$ jk\h B_z(d) -Z_{TE}\d_z B_z(d)=$$
\e =k\h(B_+ -B_-) +k_tZ_{TE}(B_+ +B_-)=0, \l{mixTE1}\f
$$ jk Z_{TM} D_z(d) - \h\d_zD_z(d)=$$
\e= k Z_{TM}(D_+ - D_-) + k_t\h(D_+ +D_-)=0, \l{mixTM1}\f
whence we can solve for two of the amplitudes as
\e B_- = \frac{k\h+k_tZ_{TE}}{k\h - k_tZ_{TE}}B_+,\f
\e D_- = \frac{kZ_{TM}+k_t\h}{kZ_{TM}-k_t\h}D_+. \f 
The fields and their derivatives at the interface $z=0$ are
\e D_z(0) =D_+  + D_-= \frac{2kZ_{TM}}{kZ_{TM}-k_t\h}D_+, \f
\e B_z(0) =B_+ + B_-= \frac{2k\h}{k\h - k_tZ_{TE}}B_+, \f
 \e \d_zD_z(0) =-jk_tD_+  + jk_tD_- = jk_t\frac{2k_t\h}{kZ_{TM}-k_t\h}D_+\f
\e \d_zB_z(0) =-jk_tB_+ + jk_tB_-= jk_t\frac{2k_tZ_{TE}}{k\h - k_tZ_{TE}}B_+.\f
Eliminating the amplitudes $D_+,B_+$ leaves us with the conditions
\e jk_t^2\h D_z(0) - kZ_{TM}\d_zD_z(0) =0, \f
\e jk_t^2Z_{TE}B_z(0) - k\h\d_zB_z(0)=0, \f
which is another set of mixed-impedance boundary conditions. Comparing these to \r{mixTE} and \r{mixTM}, we can identify the transformed parameters, denoted by primes, as
\e Z'_{TE}/\h=\h/Z_{TE},\ \ \ \ \ Z'_{TM}/\h=\h/Z_{TM}, \f
\e k'= k_t^2/k. \f
Here $\h$ is an arbitrary normalizing impedance parameter and k is a given wave number. Thus, the transformer inverts the two normalized impedance parameters. Let us check the main special cases:
\begin{itemize}
\item DB-boundary, $Z_{TE}=0, Z_{TM}=\oo$ is transformed to $Z_{TE}'=\oo, Z_{TM}'=0$, i.e., to a D'B'-boundary and vice versa.
\item PEC boundary $Z_{TE}=Z_{TM}=0$ is transformed to $Z'_{TE}=Z'_{TM}=\oo$, i.e., to a PMC boundary and vice versa.
\item The conditions of an isotropic impedance boundary $Z_{TE}=Z_{TM}=\h$, $k=k_t$ are transformed to itself.
\end{itemize}

Since a special case of the mixed-impedance conditions \r{mixTE}, \r{mixTM} of the form $\d_zD_z=0$ and $B_z=0$ can be shown to coincide with the PEC conditions \cite{AP10}, one may hope that for some value of $d$ one could actually transform the PEC boundary to the D'B' boundary. However, working with the transformation equations for general $d$ shows us that this is not possible; neither the DB nor the D'B' conditions can be obtained by transforming the PEC or PMC conditions for any choice of $d$.

\end{document}